\documentclass[twocolumn]{jpsj3}
\usepackage{txfonts}

\newcommand{\cerurhsi}{Ce(Ru$_{0.92}$Rh$_{0.08}$)$_2$Si$_2$}

\begin{document}

\title{Magnetic Field Driven Electronic Singularities through Metamagnetic Phenomena: Case of the Heavy Fermion Antiferromagnet Ce(Ru$_{0.92}$Rh$_{0.08}$)$_2$Si$_2$}

\author{
Yo~{\sc Machida}$^1$, 
Koichi~{\sc Izawa}$^1$,
Dai~{\sc Aoki}$^{2,3}$,
Georg~{\sc Knebel}$^2$,
Alexandre~{\sc Pourret}$^2$,
and
Jacques~{\sc Flouquet}$^2$\thanks{E-mail address: jacques.flouquet@cea.fr}
}

\inst{%
$^1$Department of Physics, Tokyo Institute of Technology, Meguro 152-8551, Japan\\
$^2$SPSMS, UMR-E CEA / UJF-Grenoble 1, INAC, Grenoble, F-38054, France\\
$^3$IMR, Tohoku University, Oarai, Ibaraki 311-1313, Japan
}

\date{\today}

\abst{
Thermoelectric power measurements on the heavy fermion antiferromagnet Ce(Ru$_{0.92}$Rh$_{0.08}$)$_2$Si$_2$ under magnetic field ($H$) show clearly that the antiferromagnetic state below the critical field $H_{\rm c}\sim 2.8\,{\rm T}$ can be fully decoupled from the pseudo-metamagnetic crossover at $H_{\rm m}\sim 5.8\,{\rm T}$ which occurs when the magnetization reaches a critical value. By contrast to the weak field variation of the Sommerfeld coefficient of the specific heat in the field window from $H_{\rm c}$ to $H_{\rm m}$, two electronic singularities with opposite signs are detected in the thermoelectric power. The interplay between the magnetic instability and the topological change of the Fermi surface is discussed and we argue similarities to other field instabilities in various heavy fermion compounds. 
}

\kword{metamagnetism, pseudo-metamagnetism, heavy fermion, thermoelectric power, CeRu$_2$Si$_2$}
\maketitle

\section{Introduction}
The CeRu$_2$Si$_2$ series provides one of the best  materials to clarify the quantum criticality linked to the switch from an antiferromagnetic (AF) to a paramagnetic (PM) ground state. The systems can be tuned either by pressure ($P$)~\cite{flouquet,aoki} or doping with La, Ge or Rh to the critical pressure $P_{\rm c}$ or concentration $x_c$, respectively~\cite{aoki,matsumoto,haen,aoki2}. In the AF phase, the Ising character of the magnetism leads to a clear first order metamagnetic transition at $H_{\rm c}$ from the AF to the PM state by applying the magnetic field along the $c$ axis of the tetragonal structure~\cite{flouquet,flouquet2}. 
By combination of doping and pressure tuning, the critical field $H_{\rm c}$ of the first order metamagnetic transition terminates at a quantum critical end point (QCEP) $H_{\rm c}^*\sim 4\,{\rm T}$ for the critical pressure $P_{\rm c}$ or concentration $x_c$ in La-doping case~\cite{flouquet,flouquet2,weickert}.

The paramagnetic parent Kondo-lattice compound CeRu$_2$Si$_2$ is located just at the verge of a quantum critical point (QCP)~\cite{knafo} at ambient pressure. A very sharp pseudo-metamagnetic crossover occurs at the field $H_{\rm m}$ which would join $H_{\rm c}^*$ if the lattice volume would be expanded (e.g.~by doping or negative pressure). 
Neutron scattering studies show clearly the interplay between AF and ferromagnetic (FM) fluctuations,  notably the transfer of the low energy AF excitation to the FM ones at $H_{\rm m}$ $\sim$ 7.8 T~\cite{flouquet3,sato}.
The sharp crossover at $H_{\rm m}$ is accompanied by a drastic change of the Fermi surface (FS)~\cite{lonzarich,takeshita,haoki}. This FS reconstruction is well described by assuming that the 4$f$ electron is itinerant at low field ($H < H_{\rm m}$) and localized at high field ($H$ $>$ $H_{\rm m}$)~\cite{suzuki,Aoki1993}.  However, recent synchrotron X-ray absorption spectroscopy indicates that the itinerant character of the 4$f$ electron is preserved up to high fields of 5$\times H_{\rm m}$ and a fully localized state is expected only at $H\sim 200$~T) \cite{Matsuda2012}. 
Thus the reconstruction of the FS may not be associated to a drastic change in the localization of the 4$f$ electron,  but is due to the achievement of a critical magnetic polarization of one band. 
Another scenario emphasizes that one spin-split Fermi surface is continuously suppressed giving rise to a so-called Lifshitz instability at the pseudo-metamagnetic transition~\cite{daou}.  
CeRu$_2$Si$_2$ is a compensated metal and the main features of the FS properties are:
i) below $H_{\rm m}$, the heaviest carriers are holes (for the large $\Psi$ orbit an effective mass $m^*$ up to 120~$m_0$ has been detected, $m_0$ being the free electron mass) while the electron carriers have lower $m^*$ (the highest effective mass measured for the electron FS is 20~$m_0$); 
ii) the FS reconstruction through $H_{\rm m}$ corresponds to a strong change of the electron FS centered at the $\Gamma$ point of the Brillouin zone~\cite{takeshita}.
 
Scanning the magnetic field ($H$) is a unique tool to study in detail the interplay of local Kondo type and the magnetic fluctuations with possible supplementary electronic singularities, 
such as (pseudo-)metamagnetism, if a reconstruction of the FS occurs. Under magnetic field we expect (i) a shift of the quantum criticality with respect to the zero-field states, and (ii) the decoupling between the majority and minority spin carriers.
The key question is if the pseudo-metamagnetism is driven by a FS instability or a drastic change in the nature of the magnetic correlations.
However due to the sharpness  of (i) the pseudo-metamagnetic crossover~\cite{holtmeier}, (ii) the increase of the electronic scattering at $H_{\rm m}$, and (iii) the expected concomitant increase
of the average heavy effective mass derived from thermodynamic measurements~\cite{flouquet}, 
a fine study of the FS evolution through $H_{\rm m}$ via quantum oscillations experiments is difficult.
A very nice tool to identify a drastic change in the electronic properties is the thermoelectric power (TEP referred as $S$). 
It is sensitive to the carrier type (holes leading to positive sign,
electrons to negative sign) and also to the value of the effective mass at low temperatures under the simple assumption of a free electron model.
In reality, the interpretation of TEP is rather complicated, since it depends on the energy derivatives of the density of states and the scattering time .
In a multi-carrier system, an even more complicated TEP is obtained by the sum of the contributions from different bands which are weighted by the their respective conductivity.
However, at the first approximation, the TEP is mainly dominated by the heavy band.~\cite{behnia,miyake}
This rough estimation seems to be applicable to CeRu$_2$Si$_2$.
Two decades ago, it had been recognized that the TEP of CeRu$_2$Si$_2$ varies drastically through $H_{\rm m}$ with a negative extremum just at $H_{\rm m}$~\cite{amato}. 
At low enough temperature on both sides of $H_{\rm m}$ the TEP is positive in apparent  agreement with the dHvA studies which show that the main heavy carriers are holes on both sides of $H_{\rm m}$.
Further Nernst effect measurements~\cite{behnia2} show similar drastic effects with sharp structures in contrast to the rather smooth variations detected in Hall effect experiments~\cite{kambe}. 
Recent Hall effect measurements show a small kink at $H_{\rm m}$ at very low temperatures~\cite{daou}.
The TEP measurements on CeRu$_2$Si$_2$ have been revisited down to 0.1~K and up to 12~T with the confirmation of a deep negative singularity  at $H_{\rm m}$~\cite{pfau}.

\begin{figure}[t]
\begin{center}
\includegraphics[width=1 \hsize,clip]{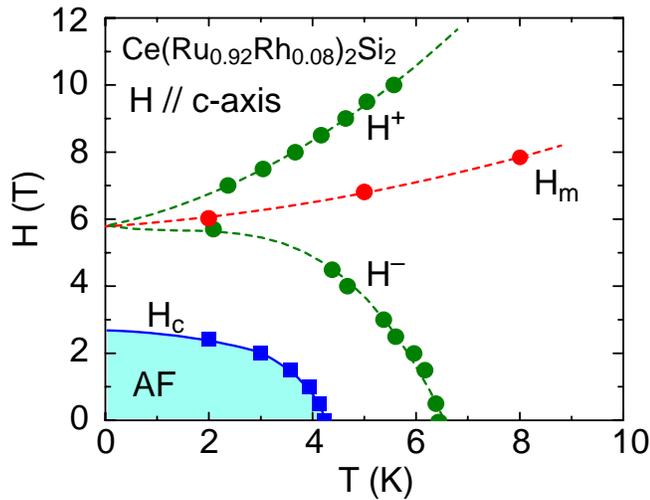}
\end{center}
\caption{\label{fig1} (Color online) ($T, H$) phase diagram of \cerurhsi\ . The antiferromagnic phase AF is limited by the  boundary $H_{\rm c}$. $H^{-}$ and $H^{+}$ represent the crossover lines determined by thermal expansion \cite{aoki2} and $H_{\rm m}$ the temperature dependence of the pseudo-metamagnetic transition.} 
\end{figure}

Doping CeRu$_2$Si$_2$ with La or Ge induces AF order and, as a consequence, $H_{\rm m}$ joins $H_{\rm c}$ at finite temperature. This inhibits the possibility to decouple at very low temperature a conventional metamagnetic transition from the AF to PM state under field and an unconventional pseudo-metamagnetic crossover which is directly related to the high magnetic polarization of the electronic band. However, recent measurements on 8$\%$ Rh doped CeRu$_2$Si$_2$ suggests that such decoupling can be realized. \cerurhsi\ is an antiferromagnet with a  N\'eel temperature of $T_{\rm N}$ = 4.2~K at zero field. The ($T,H$) phase diagram obtained by our previous experiments~\cite{aoki2} is shown in Fig.~\ref{fig1}. The AF domain  terminates at $H_{\rm c}$ = 2.8 T and a first order metamagnetic transition appears; the crossover lines $H^{-}$ and $H^{+}$ define respectively the low field PM domain with dominant AF correlation and the high polarized paramagnetic (PPM)  phase where spin up and spin down carriers start to be strongly separated~\cite{flouquet,flouquet2}.
The suggested origin of this decoupling is the change of the magnetic ordering vector and the vector of the magnetic fluctuations \cite{aoki2}. While the AF propagation vector has a transverse mode in the case of Ge and La doping with $\vec{q}=$(0.3, 0, 0) or (0.3, 0.3, 0) it becomes longitudinal $\vec{q}=$(0, 0, 0.34) for Rh doping in the AF state. Above the critical field $H_{\rm c}$ $\sim$ 2.8 T for \cerurhsi\, it is suspected that the dominant hot spot changes and a transverse mode is expected on entering in the PM regime \cite{aoki2}.
As, for this mode, the molar volume of \cerurhsi\ is smaller than that of CeRu$_2$Si$_2$, the system enters in the PM phase at $H_{\rm c}$.
Increasing further the magnetic field leads to reach a value of 0.7~$\mu_{\rm B}/{\rm Ce}$ and a pseudo-metamagnetic transition at $H_{\rm m}$ $\sim$ 5.8 T occurs~\cite{aoki2}.
Above $H_{\rm m}$, the PPM phase is quite analogous to that of the pure compound CeRu$_2$Si$_2$. 
Magnetization, specific heat and transport measurements~\cite{aoki2} underline  the striking feature that the Sommerfeld coefficient ($\gamma$) has a strong increase at $H_{\rm c}$ followed by a plateau from $H_{\rm c}$ to $H_{\rm m}$ and by a large drop above $H_{\rm m}$.
The TEP experiments reported below will show that instead of a plateau two extrema with opposite signs appear at $H_{\rm c}$ and $H_{\rm m}$, pointing out quite different electronic instabilities.

\begin{figure}[t]
\begin{center}
\includegraphics[width=0.8 \hsize,clip]{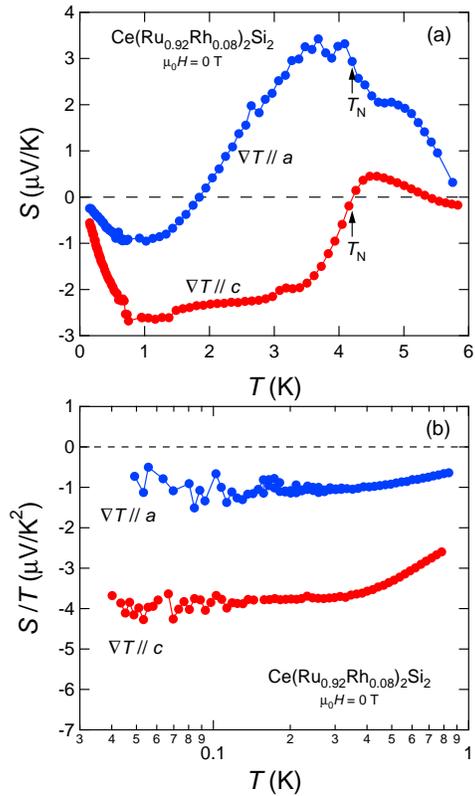}
\end{center}
\caption{\label{fig2} (Color online) (a) TEP (S) of \cerurhsi\ at $H=0$ as function of temperature for different directions of the thermal gradient $\nabla T \parallel a$ and  $\nabla T \parallel c$. (b) Temperature dependence of $S/T$ at low temperatures. $S/T$ becomes constant for $T < 0.3$~K for both configurations and is highly anisotropic.} 
\end{figure}

\section{Experimental}
We performed TEP experiments on two samples cut from a single crystal of the same batch used in previous measurements \cite{aoki, aoki2}. The thermal gradient $\nabla T$ was applied along the $a$ and $c$ axis of the tetragonal crystal, respectively. In both samples, the magnetic field was applied along the $c$ axis which is the easy magnetization axis. Experiments down to 120~mK were performed in Grenoble and down to 50~mK in Tokyo.
The interest to investigate both, the transverse ($\nabla T \parallel a \perp H$) and longitudinal configuration ($\nabla T \parallel c \parallel$ $H$) is to change the magnetoresistivity response and
thus the relative weights of the carrier scattering of the different subbands. Complementary, Hall effect experiments were also realized with $H$ $\parallel$ $c$. The magnetoresistivity has been investigated also in transverse and longitudinal configurations.

\section{Results and Discussion}
Figures~\ref{fig2}(a) and \ref{fig2}(b) show the TEP ($S$) and $S/T$ at low temperatures, with the thermal gradient $\nabla T$ applied along the $a$ and $c$ axis at $H=0$, respectively. 
For $\nabla T$ $\parallel$ $c$ the TEP jumps to a negative value at the magnetic transition at $T_{\rm N}$ and $S(T)$ remains negative down to 40~mK. 
On the other hand, for $\nabla T$ $\parallel$ $a$, $S$ increases at $T_{\rm N}$ and changes sign only at lower temperatures. 
For $T \to 0$ a smaller negative value of $S/T$ is achieved for $\nabla T \parallel a$ than for a thermal gradient applied along the $c$ axis. 
These negative limits of $S/T$ differ from the initial positive sign detected in CeRu$_2$Si$_2$ ($S/T$ near +1 $\mu$V/K$^2$ for $\nabla T$ $\parallel$ $a$) and from the general tendency reported for different Ce heavy fermion compounds.
This departure from the standard behavior points out that in this multiband material strong anisotropy occurs in the TEP. Furthermore, the contribution of each Fermi sheet (hole and electron) to the total TEP involves directly scattering due to each band through their respective electric conductivity $\sigma_i$, namely $S=\sum_i \frac{\sigma_i}{\sigma} S_i$, where $S_i$ is the contribution of TEP from each band~\cite{miyake}. Thus it is not obvious to distinguish whether the anisotropy of the total TEP is due to scattering or due to the different  Seebeck effect contributions of each band~\cite{miyake,zlatic}. 

\begin{figure}[t]
\begin{center}
\includegraphics[width=1 \hsize,clip]{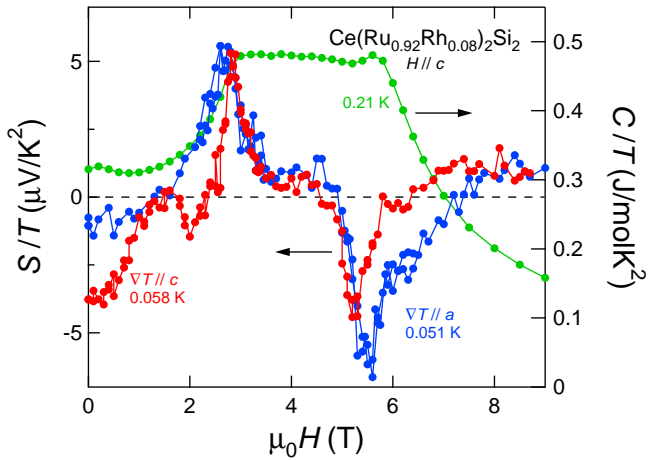}
\end{center}
\caption{\label{fig3} (Color online) Field dependence of $S/T$ of \cerurhsi\ at very low temperature for longitudinal ($\nabla T \parallel c \parallel H$) and transverse ($\nabla T \parallel a \perp H$) temperature gradient. The field is applied along the $c$ axis, which is the easy magnetization axis. For comparison the field variation of $C/T$ measured at $T=0.21$~K is shown (right scale).} 
\end{figure}

\begin{figure}[t]
\begin{center}
\includegraphics[width=1 \hsize,clip]{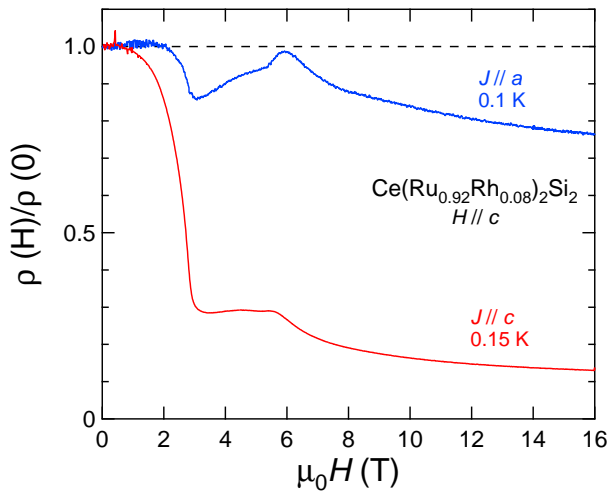}
\end{center}
\caption{\label{fig4} (Color online) Longitudinal and transverse magnetoresistivity of \cerurhsi\ at very low temperature for $H \parallel c$. } 
\end{figure}

\begin{figure}[t]
\begin{center}
\includegraphics[width=1 \hsize,clip]{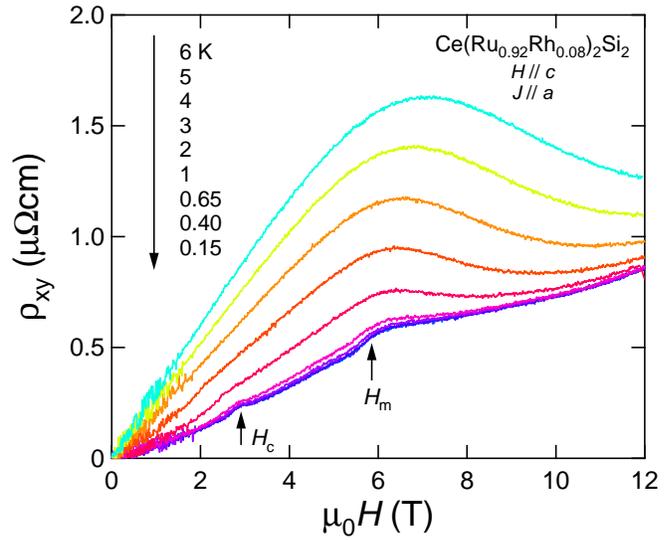}
\end{center}
\caption{\label{fig5} (Color online) Hall resistivity of \cerurhsi\ at very low temperature for $H \parallel c$.  Small anomalies appear at $H_{\rm c}$ and $H_{\rm m}$, while the Hall constant seems not to change significantly. } 
\end{figure}

The field variation of $S/T$ at very low temperatures for the two configurations is shown in Fig.~\ref{fig3} and it is compared to the field dependence of the specific heat divided by temperature $C/T$. A remarkable positive TEP anomaly appears at $H_{\rm c}$ while a negative one at $H_{\rm m}$.
The anomaly at $H_{\rm c}$ suggests a strong enhancement of the hole contribution linked to the huge increase of $\gamma$ at $H_{\rm c}$ \cite{aoki}.  The negative anomaly at $H_{\rm m}$ is quite similar to that in  CeRu$_2$Si$_2$ at its pseudo-metamagnetic transition~\cite{amato,pfau}.
For  $H < H_{\rm c}$ an additional small negative 
structure is detected close to $H \sim 2$~T  in the longitudinal configuration ($\nabla T \parallel c$) within an experimental accuracy. 
No corresponding response has been observed in the transverse configuration. 
As shown in Fig.~\ref{fig4} it may be associated to the strong decrease of the longitudinal magnetoresistivity which occur already far below $H_{\rm c}$ while for $J \parallel a$ only a weak magnetoresistance appears below  $H \sim 2$~T. 
By comparison to the anisotropic response of the TEP at zero field  the extrema observed in the field dependence of $S/T$ at $H_{\rm c}$ and $H_{\rm m}$ are quite isotropic. This points out that the $S/T$ is not governed by scattering but by 
well defined electronic instabilities. 
Small differences appear in the position and the shape of the anomaly at $H_{\rm m}$ between two configurations. 
The field $H_{\rm m}$ of the pseudo-metamagnetism is very sensitive to the sample quality. 
Tiny discrepancies in the Rh concentration between the two samples may produce a large shift of $H_{\rm m}$, while $H_{\rm c}$ shows a small shift.
 Such dependences have been pointed out previously in the study of Ce$_{1-x}$La$_x$Ru$_2$Si$_2$. \cite{flouquet}  
 Furthermore, two additional effects can be invoked:(i) the huge difference in magnetoresistivity as presented in Fig.~\ref{fig4} leading to a direct feedback on the weight by the electric conductivity, and (ii) a superstructure in the density of states reflecting the $H^{+}$ and $H^{-}$ splitting. Unfortunately, at the present stage
it is difficult to distinguish these different possibilities as doping itself leads to a large broadening of pseudo-metamagnetic crossover (near $\Delta H_{\rm m}\sim 0.5\,{\rm T}$) by quenching partly the magnetostriction at $H_{\rm m}$~\cite{aoki2}. 
Let us point out that the TEP on the pure compound CeRu$_2$Si$_2$ measured down to 1.5~K 
for the two configurations seems to indicate that the longitudinal configuration
detects $H^{-}$ while for the transverse setting $H^{+}$ could be localized~\cite{amato2}. 
Obviously, these features have to be confirmed down to very lower temperature for undoped CeRu$_2$Si$_2$. 
For comparison we performed Hall effect measurements for \cerurhsi\ as shown in Fig.~\ref{fig5}.
The Hall response $\rho_{xy}$ is dominated by the light hole charge carriers in difference to the TEP. As function of field only small anomalies appear at $H_{\rm c}$ and $H_{\rm m}$ in $\rho_{xy}$, respectively. However, there is no drastic change of Hall constant approximated through the slope $d\rho_{xy}/dH$ as can be seen in Fig.~\ref{fig5}.

The positive anomaly in the TEP at $H_{\rm c}$ shown in Fig.~\ref{fig3} is in excellent agreement with a strong enhancement of the hole  contribution at the AF/PM instability linked to the $H$ reentrance to the transverse hot spot in a PM phase. At $H_{\rm c}$, the Sommerfeld coefficient $\gamma (H_{\rm c})$ reaches 500~mJmole$^{-1}$K$^{-2}$, a value quite close to the critical value observed at $x_c$ in the Ce$_{1-x}$La$_x$Ru$_2$Si$_2$ series \cite{flouquet2}. Clearly, it appears linked to the magnetic quantum criticality of the CeRu$_2$Si$_2$ family for the transverse mode. It is not obvious if a major concomitant change of the FS occurs at $H_{\rm c}$ besides a possible slight modification of the Brillouin zone when the AF/PM borderline is crossed. 

Recently various FS studies have been reported on the CeRu$_2$Si$_2$ family to clarify the electronic states in the AF and PM regimes. The FS evolution through the critical concentration $x_c$ has been investigated by quantum oscillation measurements on CeRu$_2$(Si$_{1-x}$Ge$_x$)$_2$ series \cite{Matsumoto2011}. In this series the ground state changes from FM ($x>0.58$) to AF $(0.065 <x < 0.58)$ and finally to a PM ground state. The measurements show that the 4$f$ electron appears itinerant below the critical concentration of the transition from FM to AF order. However, up to now no complete determination of the FS in the AF phase could be realized and thus a reliable comparison with the FS of CeRu$_2$Si$_2$ cannot be achieved. However, they are in agreement with angular resolved photoemission spectroscopy (ARPES) measurements on 18 \% Ge-doped CeRu$_2$Si$_2$ ($T_{\rm N} = 8$~K) and undoped CeRu$_2$Si$_2$ (i.\/e. on both sides of $x_c$)~\cite{Okane2009} in the PM regime above $T_{\rm N}$, showing the invariance of the FS. Even for low Ce doping in LaRu$_2$Si$_2$ quantum oscillation experiments requires an itinerant treatment of the $f$ electron \cite{Matsumoto2010 ,Matsumoto2012}. However, up to now a full observation of the FS on both sides of $x_c$ (in the AF and PM state) has not been achieved. The ``common" opinion is that the magnetic quantum criticality in the CeRu$_2$Si$_2$ family
induced by La, Ge,  or Rh doping is well described by the spin density wave scenario \cite{flouquet, knafo} which is based on a continuous evolution of the FS through $x_c$.\cite{Moriya1995, Millis1993} There is no indication for an abrupt change from a ``small'' to ``large'' FS connected to quantum criticality in the CeRu$_2$Si$_2$ series in difference to the local Kondo-breakdown scenario.\cite{Si2001, Friedemann2010}

\begin{figure}[t]
\begin{center}
\includegraphics[width=1 \hsize,clip]{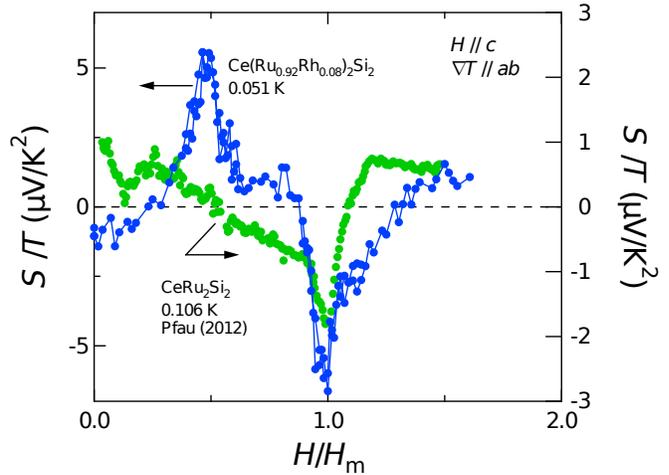}
\end{center}
\caption{\label{fig6} (Color online) Comparison of the field dependence of $S/T$ with $H \parallel c$ of CeRu$_2$Si$_2$ (data taken from Ref.~\citen{pfau}) and the doped case with 8\% Rh. } 
\end{figure}

\begin{figure}[t]
\begin{center}
\includegraphics[width=0.5 \hsize,clip]{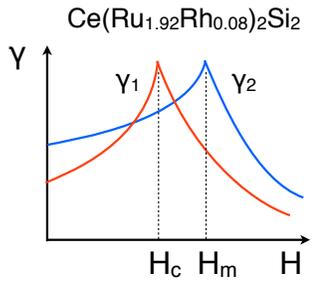}
\end{center}
\caption{\label{fig7} (Color online) Schematic plot of the field dependence of Sommerfeld coefficient $\gamma$ from different contributions as a possible explanation for the plateau of total $\gamma$-value between $H_{\rm c}$ and $H_{\rm m}$. $\gamma_1$ and $\gamma_2$ are mainly from hole and electron bands, respectively.} 
\end{figure}

In difference, the negative anomaly observed in the TEP at $H_{\rm m}$ is quite similar to that previously detected for CeRu$_2$Si$_2$ as shown the Fig.~\ref{fig6}.
Thus it marks the drastic change of the FS which is associated 
to the crossing through a critical value of the magnetic polarization under field.
Since the Sommerfeld coefficient reaches a value near $500\,{\rm mJ\,mol^{-1}K^{-2}}$ at $H_{\rm m}$ both in CeRu$_2$Si$_2$ and in \cerurhsi, the similar TEP response will be expected on the basis of a single band picture.
However, as the residual resistivity \cite{aoki, daou} and also width of the pseudo-metamagnetic crossover \cite{holtmeier} of the pure and 8\% Rh-doped crystal differ by more than one order of magnitude it is quite astonishing that only a factor 2  appears in their TEP response at $H_{\rm m}$ (see Fig.~\ref{fig6}). 
The TEP measurement proves that two different effects occur at $H_{\rm c}$ and $H_{\rm m}$ with signature from hole (at $H_{\rm c}$) and electron (at $H_{\rm m}$) charge carriers. 
Thus it supports the idea that the quasi-plateau of $\gamma (H)$ (see Fig.~\ref{fig3}) is a result of two separated effect leading to an enhancement of $\gamma$ at $H_{\rm c}$ and $H_{\rm m}$,
as schematically shown in Fig.~\ref{fig7}.
The quasi concordance of $\gamma (H_{\rm c})$ and $\gamma (H_{\rm m})$ must be connected with the observation that at $x_c$ for Ce$_{1-x}$La$_x$Ru$_2$Si$_2$ its $\gamma (x_c)$ value is comparable to $\gamma (H_{\rm m})$ in the pure compound \cite{flouquet2}.

 A key point  in the CeRu$_2$Si$_2$ series is that pseudo-metamagnetism occurs 
when the magnetization reaches a critical value~\cite{flouquet}.
It has already been proposed that the magnetic field shifts the singular peak of the density of states created by the flat band structure around (0, 0, $\pi$/c)  down the Fermi level through $H_{\rm m}$ which will lead to the strong negative TEP anomaly~\cite{miyake}. A similar conclusion may be valid on entering in the PPM phase for the  heavy fermion compounds CeCoIn$_5$~\cite{sheikin}, CeIrIn$_5$~\cite{capan} or YbRh$_2$Si$_2$~\cite{rourke}.

For CeRu$_2$Si$_2$ it is often believed  that the magnetic field will drive to an AF instability as it occurs here at $H_{\rm c}$ for Ce(Ru$_{0.92}$Rh$_{0.08}$)$_2$Si$_2$.
However, there is no indication in inelastic neutron scattering experiments of a critical slowing down of the nearly AF excitation on approaching $H_{\rm m}$. The main observation is a quasi invariance of the linewidth of the AF quasi-elastic spectrum with a field decrease of the AF signal~\cite{raymond}.
Thus clearly the pseudo-metamagnetism is mainly driven by the FS electronic instability with consequence on the nature of magnetic interaction. From the evolution of quantum oscillations  through $H_{\rm m}$ it seems clearly established that the average effective mass of the electron orbit will never exceed that of the hole orbit~\cite{takeshita, haoki}

The concept of an electronic topological transition has been developed four decades ago by Lifshitz~\cite{lifshitz} and has been discussed in detail in literature (see e.g. Ref.~\citen{varlamov}). The importance of such topological transitions for heavy fermion systems has been stressed recently~\cite{gorkov,misawa}. Experimental evidences for such electronic instabilities as function of $H$ have been claimed for several compounds like CeRu$_2$Si$_2$\cite{daou},  YbRh$_2$Si$_2$~\cite{rourke}, the FM Ising materials URhGe~\cite{yelland} and UCoGe~\cite{malone} by sweeping the field transverse to the easy magnetization axis, 
and also for the hidden order phase of URu$_2$Si$_2$~\cite{shishido,altarawneh,malone2}.
The difficulty of a quantitative treatment is to reproduce the experimental results on the basis of a fair model of the electronic band structure, as a change in one subband will affect the other bands in such multiband systems. 
This complexity may be the reason for the great diversity found in TEP anomalies among these heavy fermion compounds \cite{amato, pfau, malone, malone2}. 

Recent theoretical development on the importance of the Zeeman energy to drive Lifshitz transition can be found in refs.~\citen{kusminskiy,hackl}.

Even for a rather low average magnetic polarization due to the small magnetic field, 
the relative critical shift of the subbands accompanied with a switch to another ground state could be induced, since the drastic change of effective mass is expected near the quantum phase transition~\cite{hackl}.
A possible scenario of YbRh$_2$Si$_2$ is a cascade of Lifshitz transitions at $H_{\rm c}$~\cite{hackl,hackl2} and on entering in a PPM regime at $H_{\rm m}$~\cite{rourke}.

Quite similar phenomena occur in Yb and U heavy fermion systems as in Ce heavy fermion compounds. 
However, the difference between heavy and light carrier is far less pronounced than that for Ce compounds,
because Yb compounds have the strong localization of the 4$f$ shell and spin-orbit coupling, 
and the strong 5$f$-itinerant character appears in U compounds~\cite{flouquet4}.
The observed field induced  quantum phase transitions in these materials may be connected to electronic topological transitions. 

The difficult challenges are theoretically a sound description of the band structure and experimentally an unambiguous access to the FS of the ground state. In the CeRu$_2$Si$_2$ series, the Ising character of the magnetism opens already the possibility to observe some orbits in both, AF and PM phases \cite{Matsumoto2010}. Furthermore, the rather high N\'eel temperature (a few Kelvin for $x \sim 0.1$) may allow to clarify the FS properties through $x_c$ by ARPES. For YbRh$_2$Si$_2$ chosen often as a reference for local criticality, weakness of the AF ordering ($T_{\rm N} = 70$~mK) and  of $H_{\rm c}$ (near 60~mT for $H \parallel a$) precludes any determination of the Fermi surface by quantum oscillations.\cite{Gegenwart2002}   
On the other hand, a fully understanding of the CeRu$_2$Si$_2$ family seems very near to be achieved. 

\section{Summary}
We performed the TEP measurements down to very low temperatures under magnetic field in the heavy fermion  antiferromagnet \cerurhsi.
Two singularities of $S/T$ with opposite signs were observed at the antiferromagnetic critical field $H_{\rm c}$ and at the pseudo-metamagnetic crossover $H_{\rm m}$, which is in contrast to the plateau of Sommerfeld coefficient between $H_{\rm c}$ and $H_{\rm m}$.
We infer that it is due to the different contributions from electron and hole Fermi surface instabilities induced by the magnetic field.

\section*{Acknowledgements}
The authors thank K.~Behnia for his stimulation to revisit TEP measurements and for many discussions on the present results. Another source of motivation was given by A.~Palacio Morales with her recent data of field effects in different heavy fermion compounds.  We thank H.~ Harima and K.~Miyake for their continuous theoretical support. Finally discussions with Y.~Matsumoto give strong evidence of the possible FS status on both side of magnetic quantum critical point at $H=0$.  
This work has been supported financially by ERC starting grant (NewHeavyFermion), the French ANR projects (CORMAT, SINUS, DELICE), 
grants-in-aid from JSPS, grants-in-aid for Scientific Research on Innovative Areas ``Heavy Electrons'' (20102006) from MEXT, Global COE Program from MEXT through the Nanoscience and Quantum Physics Project of the Tokyo Institute of Technology.

\bibliographystyle{jpsj}

\end{document}